\documentclass[10pt,lettersize,journal]{IEEEtran}

\usepackage[utf8]{inputenc}
\usepackage[T1]{fontenc}
\usepackage{graphicx}
\usepackage{longtable}
\usepackage{wrapfig}
\usepackage{rotating}
\usepackage[normalem]{ulem}
\usepackage{amsmath}
\usepackage{amssymb}
\usepackage{capt-of}
\usepackage{hyperref}
\usepackage{amsmath,amssymb}
\usepackage{graphicx}
\usepackage{booktabs}
\usepackage{siunitx}
\usepackage{enumitem}
\usepackage{multirow}
\usepackage{xurl}
\usepackage{algorithm}
\usepackage{algpseudocode}
\usepackage{orcidlink}
\usepackage{algorithm}
\usepackage{algpseudocode}
\usepackage{comment}
\usepackage{amsmath,amssymb}
\usepackage{tikz}
\usetikzlibrary{arrows.meta, positioning}
\usepackage{graphicx}
\usepackage[sorting=none,giveninits=true]{biblatex}
\renewcommand\[{\begin{equation}}
\renewcommand\]{\end{equation}}
\usepackage{newtxtext}
\usepackage{newtxmath}
\usepackage{hyperref} 

\hypersetup{
 pdfauthor={},
 pdftitle={Neural Aided Adaptive Innovation-Based Invariant Kalman Filter},
 pdfkeywords={Inertial Sensors, Invariant Kalman Filter, Adaptive Techniques, Extended Kalman Filter, Inertial Navigation System, Sensor Fusion, Autonomous Underwater Vehicles},
 pdfsubject={},
 pdfcreator={},
 pdflang={English}}
\date{\today}
\begin{document}

\title{Neural Aided Adaptive Innovation-Based Invariant Kalman Filter}

\author{Barak~Diker \orcidlink{0009-0003-4551-3061} and Itzik~Klein \orcidlink{0000-0001-7846-0654}
\thanks{.}}

\maketitle

\begin{abstract}
Autonomous platforms require accurate positioning to complete their tasks. To this end, a Kalman filter-based algorithms, such as the extended Kalman filter or invariant Kalman filter, utilizing inertial and external sensor fusion are applied. To cope with real-world scenarios, adaptive noise estimation methods have been developed primarily for classical Euclidean formulations. However, these methods remain largely unexplored in the tangent Lie space, despite it provides a principled geometric framework with favorable error dynamics on Lie groups. To fill this gap, we combine invariant filtering theory with neural-aided adaptive noise estimation in real-world settings.  To this end, we derive a novel theoretical extension of classical innovation-based process noise adaptation formulated directly within the Lie-group framework. We further propose a lightweight neural network that estimates the process noise covariance parameters directly from raw inertial data. Trained entirely in a sim2real framework via domain adaptation, the network captures motion-dependent and sensor-dependent noise characteristics without requiring labeled real-world data. To examine our proposed neural-aided adaptive invariant Kalman filter, we focus on the challenging real-world scenario of autonomous underwater navigation. Experimental results demonstrate superior performance compared to existing methods in terms of position root mean square error. These results validate our sim2real pipeline and further confirm that geometric invariance significantly enhances learning-based adaptation and that adaptive noise estimation in the tangent Lie space offers a powerful mechanism for improving navigation accuracy in nonlinear autonomous platforms.
\end{abstract}


\section{Introduction}\label{sec:orgb817ba4}
Reliable state estimation is a key requirement for accurate positioning in inertial sensor fusion scenarios. In such cases, performance critically depends on the correct characterization of process and measurement noise, since the estimator must continuously balance trust between model predictions and sensor observations. The Kalman filter and its nonlinear variants remain among the most widely used frameworks for this purpose, owing to their principled probabilistic formulation and computational efficiency \cite{simon2006optimal,bar2011tracking}. However, in practical navigation scenarios, model uncertainties, nonlinear dynamics, and sensor imperfections often lead to significant deviations from idealized noise assumptions, degrading state estimation accuracy \cite{10.5555/1594745,titterton2004strapdown}.\\
Recent advances in geometric state estimation have introduced the invariant Kalman filter (IKF) framework, pioneered by Bonnabel and Barrau \cite{7523335,barrau2018invariant}. By defining estimation errors on Lie groups \cite{ye2024uncertainty} rather than in Euclidean space, the IKF often achieves linear or almost linear error dynamics for a broad class of systems \cite{islam2026error,bonnable2009invariant,potokar2021invariant,barrau2017three,hartley2020contact,bonnabel2007left}, and can enforce geometrical constraint naturally~\cite{barrau2019extended}. A central theoretical result, known as the log-linear property of the error \cite{7523335}, shows that under specific structural conditions, the nonlinear estimation error evolves according to a linear system in the associated Lie algebra space. This property enables Kalman filtering directly in the tangent space while preserving the geometric structure of the underlying system \cite{barrau:tel-01247723}. Moreover, invariant error representations naturally encode complex, non-Gaussian distributions from the original state space as Gaussian distributions within the tangent space, including highly nonlinear, banana-shaped uncertainties \cite{barrau:tel-01247723,long2013banana}. \\
Despite these advantages, a fundamental challenge remains: the tuning and adaptation of process noise covariance matrices to suit diverse real-world scenarios. In classical Kalman filtering \cite{paul2005fundamentals}, numerous adaptive techniques have been proposed to estimate noise statistics online, including innovation-based methods and covariance matching approaches \cite{mohamed1999adaptive,Ding_Wang_Rizos_Kinlyside_2007,11361209,liu2018innovative}, and open-loop neural-based adaptive Kalman filters that adapt the process covariance matrix online \cite{solodar2024vio, 11361209}. However, these methods are formulated in Euclidean error spaces and do not directly extend to Lie-group-based estimators. Consequently, there is a lack of systematic investigation into adaptive noise estimation in the tangent Lie space, where invariant filters operate. \\
Only limited progress has been made in this direction. Kim et al. \cite{kim2025adaptive}  proposed an adaptive invariant filtering scheme that switches between the right and the left invariant Kalman filter based on predefined threshold of the trace of the covariance. Another closely related work by Brossard et al. \cite{brossard2020denoising} learns process noise statistics using neural networks, but the approach is evaluated primarily in simulation and not integrated into real-world inertial navigation pipelines. To the best of our knowledge, no prior work has systematically studied adaptive process noise estimation for invariant Kalman filters using real underwater navigation data and direct comparison with established adaptive EKF techniques. \\
This paper fills this gap by combining invariant filtering theory with neural-aided adaptive noise estimation in real-world scenarios. Building on the theoretical foundations of invariant Kalman filtering \cite{7523335} and learning-based noise estimation \cite{brossard2020denoising}, we derive the theory for adaptive process noise covariance estimation in the tangent Lie space. 
We design a simple yet effective neural network architecture for the regression task, which processes a set of inertial measurements to adaptively output the diagonal elements of the process noise covariance matrix. Since ground-truth process noise is inherently latent in real-world data, we adopt a simulation-to-real (sim2real) training approach to leverage perfect ground-truth information. Additionally, for practical reasons, using a simulated dataset eliminates the need for labor-intensive data collection. \\
To demonstrated our proposed neural-aided adaptive IKF, we focus on the challenging real-world scenario of underwater navigation. Specifically, on autonomous underwater vehicle (AUV) navigation with fusion between Doppler velocity log (DVL) data and inertial sensors \cite{cohen2024seamless,cohen2024inertial,potokar2021invariant,du2024novel,li2025nonlinear,11355492}.%
\\
The main contributions of this paper are as follows:
\begin{enumerate}
\item \textbf{Invariant adaptive process noise formulation:} 

Derivation of a novel theoretical extension of classical innovation-based process noise adaptation formulated directly within the Lie-group framework. The proposed adaptation law is derived using the invariant innovation definition and is fully consistent with the error dynamics of the IKF. Unlike conventional Euclidean formulations, the proposed method preserves the geometric structure of invariant filtering while remaining compatible with established innovation-based covariance adaptation principles, thereby bridging classical adaptive filtering theory and invariant state estimation.
\item \textbf{Neural-aided process noise identification:} Development of a lightweight and computationally efficient neural network that estimates the process noise covariance parameters directly from raw inertial data. Trained entirely in a sim2real framework via domain adaptation, the network captures motion-dependent and sensor-dependent noise characteristics without requiring labeled real-world data, enabling adaptive uncertainty modeling beyond fixed analytical noise assumptions.
\item \textbf{Hybrid invariant adaptive framework:} Formation of a unified hybrid adaptive strategy that systematically combines invariant innovation-based covariance adaptation with neural-aided noise estimation. The resulting framework leverages complementary strengths: model-based innovation statistics ensure consistency and correction of residual modeling errors, while the neural network provides responsive, data-driven noise adaptation. This integration yields improved estimation accuracy and robustness while preserving the theoretical guarantees and structural properties of the IKF.
\end{enumerate}
To demonstrate our proposed approach, we employ the A-KIT dataset \cite{cohen2025adaptive}, consisting of 80 minutes of real recorded data in different AUV missions. A rigorous comparison between classical EKF, adaptive EKF, invariant EKF (IEKF), and neural adaptive invariant filtering approaches is made. We show that our approach achieves superior performance compared to the other methods in terms of the position root mean square error. Thus, we provide a practical and scalable solution for high-precision underwater navigation. \\
The remainder of this paper is organized as follows. Section~\ref{sec:org66cbe53} formulates the navigation problem and introduces the Lie-group-based system model. Section~\ref{sec:orgde6f5a2} presents our proposed neural-aided adaptive innovation-based IKF framework and training methodology. Section~\ref{sec:orgc0ce113} provides a comprehensive experimental evaluation and comparative analysis of the proposed approach against baseline methods. Finally, Section~\ref{sec:org5d74686} concludes the paper and discusses future research directions.
\section{Problem Formulation}
\label{sec:org66cbe53}
\subsection{Inertial Navigation}
\label{sec:org5ea6611}
The continuous-time navigation dynamics, expressed in the earth centered earth frame (ECEF), are given by \cite{10.5555/1594745}:
\begin{equation}
\begin{aligned}
&\dot{\boldsymbol{p}}_{eb}^{e} = \boldsymbol{v}_{eb}^{e} \\
&\dot{\boldsymbol{v}}_{eb}^{e} = \mathbf{C}_{b}^{e} \boldsymbol{f}_{ib}^{b}
- 2(\boldsymbol{\omega}_{ie}^{e})^{\wedge} \boldsymbol{v}_{eb}^{e}
+ \boldsymbol{g}_{ib}^{e} \\
&\dot{\mathbf{C}}_{b}^{e} = \mathbf{C}_{b}^{e}(\boldsymbol{\omega}_{ib}^{b})^{\wedge}
- (\boldsymbol{\omega}_{ie}^{e})^{\wedge} \mathbf{C}_{b}^{e}
\end{aligned}
\end{equation}
where \(\boldsymbol{\omega_{ib}^{b}}\) and \(\boldsymbol{f}_{ib}^{b}\) denote the measured angular velocity and specific force in the body frame, \(\boldsymbol{\omega_{ie}^{e}}\) is the Earth rotation rate expressed in the navigation frame, \(\boldsymbol{g}_{ib}^{e}\) is the gravity vector, \(C_{b}^{e}\in SO(3)\) is the rotation matrix from body frame to ECEF frame, \(v^{e}\) is the velocity of body with respect to ECEF frame expressed in ECEF frame, and \(p^{e}\in \mathbb{R}^{3}\) describes the position of body with respect to ECEF frame expressed in ECEF frame. The operator \((\cdot)^{\wedge}\) denotes a mapping between vector to a the lie algebra set \(\mathfrak{so}(3)\). 
\subsection{Invariant Kalman Filter Formulation}
\label{sec:org3666251}

Recent developments in invariant observer theory advocate for defining estimation errors in a manner that respects the geometric structure of the state space. Rather than employing the additive errors typical of the error state EKF, the IKF defines errors directly on the Lie group. \\
The attitude error is defined as:
\begin{equation}
\hat{\mathbf{C}}_{b}^{e} \mathbf{C}_{e}^{b} = \exp(\boldsymbol{\phi^{e}}) \approx I + (\mathbf{\phi^{e}})^\wedge
\end{equation}
while the velocity and position errors are defined as:
\begin{equation}
\begin{aligned}
& \boldsymbol{\eta_{v}^{R}} := \mathbf{J} \boldsymbol{\rho_{v}^{e}} = \hat{\boldsymbol{v}}_{eb}^{e} - \hat{\mathbf{C}}_{b}^{e} \mathbf{C}_{e}^{b} \boldsymbol{v}_{eb}^{e}
\\
& \boldsymbol{\eta_{p}^{R}} := \mathbf{J} \boldsymbol{\rho_{p}^{e}} = \hat{\boldsymbol{p}}_{eb}^{e} - \hat{\mathbf{C}}_{b}^{e} \mathbf{C}_{e}^{b} \boldsymbol{p}_{eb}^{e}
\end{aligned}
\end{equation}
where \(\mathbf{J}_r(\boldsymbol{\phi})\) is the right Jacobian of \(\mathrm{SO}(3)\) associated with the exponential map, mapping perturbations in the Lie algebra to right-multiplicative perturbations on the manifold, and is given by:
\begin{equation}
\begin{aligned}
\mathbf{J}_r(\boldsymbol{\phi^{e}})
&=
\mathbf{I}
-\frac{1-\cos\|\boldsymbol{\phi^{e}}\|}{\|\boldsymbol{\phi^{e}}\|^2}\,(\boldsymbol{\phi^{e}})^{\wedge} \\
&+\frac{\|\boldsymbol{\phi^{e}}\|-\sin\|\boldsymbol{\phi^{e}}\|}{\|\boldsymbol{\phi^{e}}\|^3}\,((\boldsymbol{\phi^{e}})^{\wedge})^2
\end{aligned}
\end{equation}

Although these error definitions may appear more involved than their classical counterparts, they lead to error dynamics that are nearly autonomous, meaning that they depend weakly on the estimated state. \\
With these definitions, the invariant error dynamics take the linearized form 
\begin{equation}
\label{err-dynamics-lie}
\frac{d}{dt}
\begin{bmatrix}
\boldsymbol{\phi^{e}} \\
\mathbf{J}\boldsymbol{\rho_{v}^{e}} \\
\mathbf{J}\boldsymbol{\rho_{p}^{e}} \\
\delta \boldsymbol{b}_{g} \\
\delta \boldsymbol{b}_{a}
\end{bmatrix}
=
\mathbf{F}
\begin{bmatrix}
\boldsymbol{\phi^{e}} \\
\mathbf{J}\boldsymbol{\rho_{v}^{e}} \\
\mathbf{J}\boldsymbol{\rho_{p}^{e}} \\
\delta \boldsymbol{b}_{g} \\
\delta \boldsymbol{b}_{a}
\end{bmatrix}
+
\mathbf{G}
\begin{bmatrix}
\boldsymbol{\omega_{\psi}} \\
\boldsymbol{\omega_{a}} \\
\boldsymbol{\omega_{b_{\psi}}} \\
\boldsymbol{\omega_{b_{a}}}
\end{bmatrix}
\end{equation}

The system and noise matrices are given by \cite{luo2021se2}:
\begin{equation}
\label{F-IKF}
\resizebox{\columnwidth}{!}{$
\mathbf{F} =
\begin{bmatrix}
-(\boldsymbol{\omega_{ie}^{e}})^{\wedge} & 0 & 0 & \hat{\mathbf{C}}_{b}^{e} & 0 \\
-(\boldsymbol{v}_{eb}^{e})^{\wedge}(\boldsymbol{\omega_{ie}^{e}})^{\wedge} + (\boldsymbol{g}_{ib}^{e})^{\wedge} & -2(\boldsymbol{\omega_{ie}^{e}})^{\wedge} & 0 & (\hat{\boldsymbol{v}}_{eb}^{e})^{\wedge} \mathbf{C}_{b}^{e} & \hat{\mathbf{C}}_{b}^{e} \\
-(\boldsymbol{p}_{eb}^{e})^{\wedge}(\boldsymbol{\omega_{ie}^{e}})^{\wedge} & I & 0 & (\hat{\boldsymbol{p}}_{eb}^{e})^{\wedge} \mathbf{C}_{b}^{e} & 0 \\
0 & 0 & 0 & 0 & 0 \\
0 & 0 & 0 & 0 & 0
\end{bmatrix}
$}
\end{equation}
and 
\begin{equation}
\label{G-IKF}
\mathbf{G} =
\begin{bmatrix}
\mathbf{C}_{b}^{e} & 0 & 0 & 0 \\
(\boldsymbol{v}_{eb}^{e})^{\wedge} \mathbf{C}_{b}^{e} & \mathbf{C}_{b}^{e} & 0 & 0 \\
(\boldsymbol{p}_{eb}^{e})^{\wedge} \mathbf{C}_{b}^{e} & 0 & 0 & 0 \\
0 & 0 & I & 0 \\
0 & 0 & 0 & I
\end{bmatrix}
\end{equation}
Compared to the  error state EKF, the IKF system matrix exhibits reduced dependence on the estimated orientation and stronger dependence on position and velocity states. This structural property underpins the improved robustness of the IKF, particularly when combined with adaptive or learning-based process noise modeling techniques.
\subsection{Measurement Model for IKF}
\label{sec:orgcc51617}
The IKF error \cite{barrau:tel-01247723} is defined as
\begin{equation}
\label{error-def}
\mathbf{\eta}=\mathcal{\hat{X}}\mathcal{X}^{-1}=expm(\mathbf{\zeta^{\wedge}})
\end{equation}
where \(expm(\mathbf{A})\triangleq\sum_{n=0}^{\infty}\frac{1}{n!}\mathbf{A}^{n}\) is the standard matrix exponential function
\begin{equation}
\label{x-variable}
\mathcal{X} =
\begin{bmatrix}
\mathbf{C}_{b}^{e} & \boldsymbol{v}_{eb}^{e} & \boldsymbol{p}_{eb}^{e} \\
0 & 1 & 0 \\
0 & 0 & 1
\end{bmatrix}
\end{equation}
and \(\zeta^{\wedge} \in \mathbb{R}^{9}\) is the new error vector mapped by extension of the \((.)^{\wedge}\) function such that
\begin{equation}
\label{vec2lie}
(\cdot)^{\wedge} : \xi \in \mathbb{R}^{9} \mapsto
\begin{bmatrix}
0 & -\xi_3 & \xi_2 & \xi_4 & \xi_7 \\
\xi_3 & 0 & -\xi_1 & \xi_5 & \xi_8 \\
-\xi_2 & \xi_1 & 0 & \xi_6 & \xi_9 \\
0 & 0 & 0 & 0 & 0 \\
0 & 0 & 0 & 0 & 0 \\
\end{bmatrix}
\end{equation}
%
The DVL measurement in IKF formulation is
\begin{equation}
\label{new-measurement-y}
\boldsymbol{Y} = \mathcal{X}^{-1}\boldsymbol{b} + \boldsymbol{V} =  \begin{bmatrix} \boldsymbol{v}_{eb}^{b} & -1 & 0 \end{bmatrix}^{T}
\end{equation}
where
\begin{equation}
\label{b-vector}
\boldsymbol{b} := \begin{bmatrix} 0 & 0 & 0 & -1 & 0 \end{bmatrix}^{T}
\end{equation}

The corresponding innovation is
%
\begin{equation}
\label{innov-equ}
\boldsymbol{r_{innov} }= \mathbf{H}\boldsymbol{\xi} + \hat{\mathcal{X}}\boldsymbol{V}
\end{equation}
where
\begin{align}
\mathbf{H} &= 
\begin{bmatrix}
0_{3\times3} & -I_{3\times3} & 0_{3\times3} & 0_{3\times3} & 0_{3\times3}
\end{bmatrix}^{T}, 
\label{H-IKF} \\
\boldsymbol{\xi} &= 
\begin{bmatrix}
\boldsymbol{\xi}_{\theta} &
\boldsymbol{\xi}_{v} &
\boldsymbol{\xi}_{p} &
\delta \boldsymbol{b}_{g} &
\delta \boldsymbol{b}_{a}
\end{bmatrix}^{T}
\in \mathbb{R}^{9}
\label{xi}
\end{align}
and
\begin{equation}
\Pi = \begin{bmatrix} I_{3\times 3 } &  0_{3\times 2} \end{bmatrix}
\end{equation}
In order to justify the usage of the Kalman filter equations, the relation between the postrior error and the a-prior error should comply to \cite{paul2005fundamentals,10.5555/1594745,grove2013principles} 
\begin{equation}\label{eq:etap}
\boldsymbol{\xi^{+} }= (I-\mathbf{KH})\boldsymbol{ \xi} -\mathbf{K}\boldsymbol{V} 
\end{equation}
where \(K\) is the Kalman gain\\
The right invariant error is \cite{barrau:tel-01247723}:
\begin{align}
\label{error-def-measure}
\mathbf{\eta }&= \hat{\mathcal{X}}( \hat{\mathcal{X}}^{+})^{-1} \approx expm(\zeta^{\wedge}_{1:9}) 
\end{align}
Given (\ref{error-def-measure}), (\ref{innov-equ}), \(\boldsymbol{\zeta} = \mathbf{K}\times  \boldsymbol{r_{innov}}\), and using the approximation \(expm(\zeta^\wedge_{1:9})\approx I + \zeta^{\wedge}_{1:9}\), \eqref{eq:etap} is rewritten as:
\begin{equation}
\label{derived-H}
\boldsymbol{\xi^{+} }= (I-\mathbf{KH})\boldsymbol{ \xi} -\mathbf{K}\mathcal{X}\boldsymbol{V}_{aug} 
\end{equation}
where
\begin{equation} 
\label{v-extend}
\boldsymbol{V}_{aug}=\begin{bmatrix} \boldsymbol{V} & 0_{1\times 2} \end{bmatrix}^{\top}
\end{equation}
justifying the usage of the Kalman filter equations. 
\subsection{Right-Invariant Kalman Filter}
\label{sec:org5066bf3}
The right-invariant error is defined as:
\[
\boldsymbol{\eta} \triangleq \hat{\mathcal{X}}\mathcal{X}^{-1}
\qquad
\boldsymbol{\eta}^{+} \triangleq \hat{\mathcal{X}}^{+}\mathcal{X}^{-1}
\]
where $\mathcal{X}$ as defined in \eqref{x-variable}
is the GT of the state, \(\hat{\mathcal{X}}\) is the estimated state, and \(\hat{\mathcal{X}}^{+}\) is the posterior state after the update phase.
By the log-linear property \cite{7523335} the nonlinear error \(\boldsymbol{\eta}\) at all times satisfies the relation
\begin{equation}
\boldsymbol{\eta} \simeq expm(\boldsymbol{\xi}^{\wedge})
\end{equation}
even when the nonlinear error is arbitrarily large. \\
The correction is applied through a group action
\begin{align}
\boldsymbol{\zeta} 
&= \mathbf{K}\,\boldsymbol{r}_{\text{innov}} \\
\hat{\mathcal{X}}(\hat{\mathcal{X}}^{+})^{-1} 
&= \operatorname{expm}\!\left(\boldsymbol{\zeta}^{\wedge}_{1:9}\right) \\
\hat{b}^{+}_{g} 
&= \hat{b}_{g} + \zeta_{10:12} \\
\hat{b}^{+}_{a} 
&= \hat{b}_{a} + \zeta_{13:15}
\end{align}
which implies (first-order) the invariant error update
\[
\exp(-\boldsymbol{\zeta}^{\wedge}_{1:9})\,\boldsymbol{\eta} = \boldsymbol{\eta}^{+}
\Rightarrow\boldsymbol{\xi}^{+} = (\mathbf{I}-\mathbf{K}\mathbf{H})\boldsymbol{\xi}-\mathbf{K}\,\mathcal{X}\,\boldsymbol{V}_{aug}
\]
where \(\boldsymbol{V}_{aug}\) is defined in (\ref{v-extend}).
\subsection{Propagation (Prediction)}
\label{sec:orgbafcebe}

The invariant error coordinates follow a linearized dynamics of the form
\[
\dot{\boldsymbol{\xi}} = \mathbf{F}\boldsymbol{\xi} + \mathbf{G}\mathbf{w}
\qquad
\mathbf{w}\sim\mathcal{N}(\mathbf{0},\mathbf{Q})
\]
with discrete-time transition matrix
\[
\mathbf{\Phi} \triangleq expm(\mathbf{F}\Delta t)
\]
where $\mathbf{F}$ is defined in (\ref{F-IKF}).\\
The error state covariance propagation is therefore
\begin{equation}
\label{propagate-IKF}
\mathbf{P}_{n|n-1} = \mathbf{\Phi}\,\mathbf{P}_{n-1|n-1}\,\mathbf{\Phi}^{\top} + \mathbf{Q}_{d}
\end{equation}
where \(\mathbf{G}\) is as defined in (\ref{G-IKF})
and the discrete process noise covariance is: \
\[
\mathbf{Q}_{d} \approx \int_{0}^{\Delta t}\mathbf{\Phi}(\tau)\mathbf{G}\mathbf{Q}\mathbf{G}^{\top}\mathbf{\Phi}(\tau)^{\top}d\tau
\]
%
\subsection{Update (Correction)}
\label{sec:org0e1080d}
The innovation is defined at \eqref{innov-equ} and its corresponding covariance are derived using \eqref{derived-H} as \cite{7523335}:
\begin{align}
&\boldsymbol{r}_{\text{innov}} 
\triangleq \Pi(\hat{\mathcal{X}}^{-1}\boldsymbol{y}- \boldsymbol{b}) \\
&\mathbf{S} 
= \mathbf{H}\mathbf{P}_{n|n-1}\mathbf{H}^{\top} 
+ \Pi (\mathcal{X}\mathbf{R}_{\text{aug}}\mathcal{X}^{\top} )\Pi^{\top}
\end{align}
where $\mathbf{H}$ is defined in (\ref{H-IKF}) and the measurement noise covariance is:
\begin{equation}
\label{R-extend}
\mathbf{R}_{aug} = 
\begin{pmatrix}
\mathbf{R} & 0_{3\times 2} \\
0_{2\times 3} & 0_{2\times 2}
\end{pmatrix}
\end{equation}
The Kalman gain is calculated by:
\[
\mathbf{K} = \mathbf{P}_{n|n-1}\mathbf{H}^{\top}\mathbf{S}^{-1}.
\]
The updated error state covariance  is
\begin{equation}
\mathbf{P}_{n|n} = (\mathbf{I}-\mathbf{K}\mathbf{H})\mathbf{P}_{n|n-1}
\end{equation}
%
Finally, the corrected estimate is obtained by retracting the correction onto the group
\begin{equation}
\label{correction-on-IKF}
\hat{\mathcal{X}}^{+} = expm\!\left((-\mathbf{K}\boldsymbol{r_{innov}})^{\wedge}\right)\hat{\mathcal{X}}
\end{equation}
\section{Proposed approach} 
\label{sec:orgde6f5a2}
\begin{figure*}
\centering
\includegraphics[width=\textwidth]{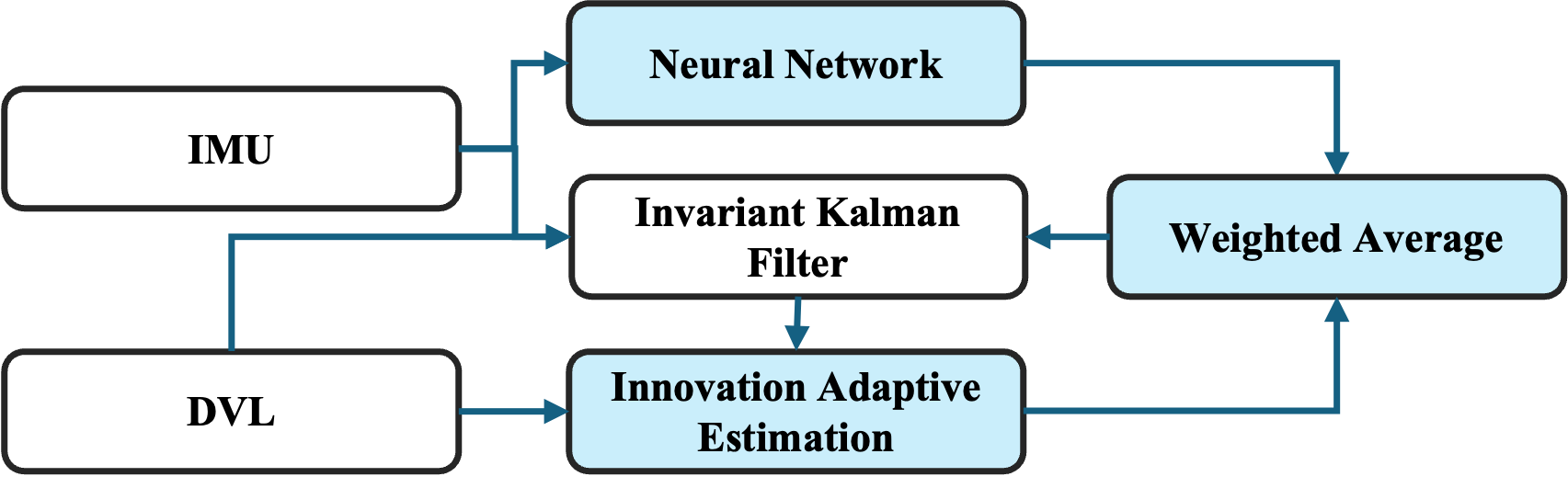}
\caption{Block diagram of the proposed neural-aided adaptive innovation-based IKF for inertial/DVL fusion.}
\label{fig:nn-architecture}
\end{figure*}

We propose a neural-aided adaptive process noise modeling framework for
inertial navigation based on the IKF, in which the
process noise covariance is tuned online to improve state estimation
accuracy and to better reflect the current sensor noise characteristics and specific motion conditions. \\
To achieve this, we combine two complementary sources of information for process noise adaptation. First, we extend the traditional innovation-based covariance adaptation approach to the IKF framework. Classical innovation-based methods adjust the process noise covariance by analyzing the statistical properties of recent filter innovations in order to compensate for model mismatch and unmodeled disturbances. In our formulation, this concept is adapted to the invariant filtering setting, where the innovation structure and error dynamics evolve on a Lie group and the formula derived at \cite{mohamed1999adaptive} isn't compatible. Second, we employ a simple yet effective neural network that receives a sliding window of raw inertial data and outputs a set of parameters characterizing the current gyroscope and accelerometer noise intensities. These parameters define a diagonal continuous-time process noise covariance matrix that is incorporated into the filter dynamics, providing a lightweight and responsive mechanism for motion-dependent noise adaptation. \\
The final process noise covariance used in the IKF propagation step is obtained
by an appropriate combination (averaging or weighted sum) of the
neural-network-based estimate and the innovation-based covariance estimate,
leveraging the strengths of both model-based and data-driven adaptation. This
hybrid strategy enables robust and responsive noise tuning while preserving
filter stability.
For training, the neural network is developed entirely in a sim2real 
framework. The motivation of using a sim2real approach is twofold:
\begin{enumerate}
\item The actual process noise is unknown in real world data.
\item Simulative data offer perfect ground truth information and eliminates the need for labor-intensive data collection.
\end{enumerate}
To this end, a diverse set of synthetic inertial data is generated through domain randomization over a wide range of sensor noise levels, bias instability
parameters, vehicle motion profiles, and environmental disturbances. The
network is trained to infer the underlying process noise characteristics from
these simulated measurements, allowing it to generalize to real-world
conditions without requiring labeled real data.
\subsection{Adaptive Hybrid Invariant Kalman Filter}
\label{sec:org38e903f}

In a standard EKF, the innovation is defined in a Euclidean sense as the difference between the measured and predicted outputs (\ref{innov-eskf}), namely
\begin{equation}
\label{innov-eskf}
\delta \boldsymbol{y} = \boldsymbol{y} - \hat{\boldsymbol{y}} 
\end{equation}
This Euclidean innovation is the basis for the adaptive process noise tuning approach proposed in \cite{mohamed1999adaptive}. \\
In contrast, in the IKF framework considered in this work, the innovation is not naturally defined as a Euclidean difference. Instead, it is expressed on the underlying Lie group through a projection map \(\Pi(\cdot)\), leading to the invariant innovation (\ref{innov-equ})
\begin{align}\label{innov-ikf}
 \boldsymbol{v}_{i}:=\boldsymbol{r}_{\text{innov}} 
&= \Pi\left( \hat{\mathcal{X}} \boldsymbol{Y} - \boldsymbol{b} \right)
\in \mathbb{R}^{3}
\end{align}
where the projection map in the case of velocity aiding expressed in the body frame is defined as:
\begin{equation}
\label{projection-map}
\Pi = \begin{bmatrix} I_{3\times 3 } &  0_{3\times 2} \end{bmatrix}
\end{equation}
Consequently, the direct application of the innovation-based adaptation scheme in \cite{mohamed1999adaptive}, which relies explicitly on a Euclidean innovation of the form (\ref{innov-eskf}), does not immediately carry over to the invariant setting. \\
The adapted process noise covariance matrix at time step \(k\) is then computed as:
\begin{equation}
\label{adaptive-Q}
\hat{\mathbf{Q}}_{r_{innov}}^{k}
= \frac{1}{N} \sum_{j=k-N+1}^{k}\mathbf{K}_{j}\Pi(\mathcal{X}_{j}\boldsymbol{y}_{j}-\boldsymbol{b})[\Pi(\mathcal{X}_{j}\boldsymbol{y}_{j}-\boldsymbol{b})]^{t}\mathbf{K}_{j}^{t}
\end{equation}
where \(\mathbf{K}_n\) denotes the Kalman gain at time step \(n\). \\
This formulation preserves the spirit of the adaptive approach in \cite{mohamed1999adaptive}, while ensuring consistency with the geometric structure of the invariant Kalman filter. \\
To enhance robustness and maintain filter consistency, the adaptive invariant innovation-based estimate of the process noise is combined with a lightweight, open-loop learning-based neural network that estimates the underlying inertial process noise parameters from raw inertial measurements. Let \(\mathbf{Q}_{r_{innov}}\) denote the innovation-based estimate of the process noise covariance in (\ref{adaptive-Q}), which is computed online during filter operation.
Accordingly, the covariance propagation (\ref{propagate-IKF}) by using (\ref{F-IKF}) (\ref{G-IKF}) is modified as
\begin{equation}
\label{hybrid-adaptive-Q}
\mathbf{P}_{k+1} = \mathbf{F}_k \mathbf{P}_k \mathbf{F}_k^\top + \lambda \mathbf{G}_k \mathbf{Q}_{\text{NN}}\mathbf{G}_k^\top + (1-\lambda) \mathbf{Q}_{r_{innov}}^{k}
\end{equation}
where \(\lambda \in [0,1]\) is a blending parameter that balances the contribution of the learning-based and innovation-based noise models, \(\mathbf{Q}_{r_{innov}}\) is the invariant innovation-based noise model (\ref{adaptive-Q}), \(\mathbf{Q}_{\text{NN}}\) is the learning-based neural covariance (\ref{Q-NN}), \(\mathbf{F}_k\) denotes the discrete-time state-transition matrix, and \(\mathbf{G}_k\) is the noise shaping matrix that maps continuous-time process noise to discrete time. \\
The blending parameter $\lambda$ controls the relative influence between the neural-aided covariance prediction and the model-based innovation adaptive correction. In this work, the value $\lambda = 0.6$ is used, giving slightly higher weight to the neural network-predicted covariance while still preserving a significant contribution from the innovation-based adaptive term. This choice was determined empirically and provides a good compromise between learning-based modeling capability and filter robustness. In particular, it prevents the estimator from relying entirely on the neural network predictions while still allowing it to capture complex noise characteristics that may not be fully represented by the analytical innovation-based approach. \\
This hybrid formulation integrates global, motion-aware noise adaptation provided by the neural network with local, data-driven corrections from the innovation-based estimate.
\subsection{Neural Network Architecture}
\label{sec:org80766c3}
The neural network architecture used for process noise estimation is described in Table \ref{fig:neuralnetwork1} . The input to the network is a window of inertial measurements of shape (B,C,W), where 
B denotes the batch size, 
C=6 corresponds to the sensor axes, and 
W=100 is the window length.
The first stage consists of a stack of three one-dimensional convolutional layers operating along the temporal dimension. These layers use kernel size 5 with symmetric padding, and progressively increase the feature dimension from 32 to 64 and 128 channels. Each convolution is followed by batch normalization and a LeakyReLU activation \cite{maas2013rectifier}, which improves gradient flow and robustness to varying signal magnitudes.
The resulting feature maps are aggregated using temporal mean pooling, yielding a fixed-dimensional representation independent of the window length. This pooled feature vector is then passed through a fully connected regression head composed of two hidden layers with 64 units each, interleaved with LeakyReLU activations and dropout regularization.
The final layer outputs a six-dimensional vector corresponding to the noise scale parameters. A softplus activation \cite{rumelhart1985learning} function is applied to enforce positivity, followed by a small constant offset to ensure numerical stability. \\
The learned process noise covariance matrix is defined as a diagonal matrix
\(Q_{\mathrm{NN}} \in \mathbb{R}^{12 \times 12}\), corresponding to the gyroscope,
accelerometer, gyroscope bias, and accelerometer bias noise terms. The structure
of this matrix is given by
\begin{equation}
\label{Q-NN}
\mathbf{Q}_{\mathrm{NN}} =
\begin{aligned}
\mathrm{diag}\Big(
q_{\omega_x},\, q_{\omega_y},\, q_{\omega_z},\,
q_{a_x},\, q_{a_y},\, q_{a_z},\, \\
q_{b_{\omega_{x}}}, q_{b_{\omega_{y}}}, q_{b_{\omega_{z}}},\,
q_{b_{a_{x}}}, q_{b_{a_{y}}}, q_{b_{a_{z}}}
\Big)
\end{aligned}
\end{equation}
where \(q_{\omega_i}\) and \(q_{a_i}\) represent the process noise variances
associated with the gyroscope and accelerometer measurements along the three
body axes, respectively. Similarly, \(q_{b_{\omega_{i}}}\) and \(q_{b_{a_{i}}}\)
denote the process noise variances corresponding to the gyroscope and
accelerometer bias dynamics for each axis. \\
Let
\[
\mathbf{z}_{\mathrm{IMU}}^{(k:k+T)} 
= \{ \mathbf{z}_{\mathrm{IMU}}(k),\mathbf{z}_{\mathrm{IMU}}(k+1),\dots,
\mathbf{z}_{\mathrm{IMU}}(k+T) \}
\]
denote a sliding window of raw inertial measurements of length \(T\). The neural
network maps this window to a six-dimensional vector of process noise
parameters
\begin{equation}
\label{neural-network-input}
\begin{split}
\mathbf{q}_{\text{NN}}
&= f_{\theta}\!\left(\mathbf{z}_{\mathrm{IMU}}^{(k:k+T)}\right) \\
&=
\begin{bmatrix}
q_{\omega_x} & q_{\omega_y} & q_{\omega_z} &
q_{a_x} & q_{a_y} & q_{a_z}
\end{bmatrix}^{\!\top}
\end{split}
\end{equation}
where \(f_{\theta}(\cdot)\) denotes the neural network with parameters \(\theta\).
The elements of \(\boldsymbol{q}_{\text{NN}}\) correspond to the process noise
variances associated with the gyroscope and accelerometer measurements along
the three body axes, and directly determine the first six diagonal entries of
\(\mathbf{Q}_{\mathrm{NN}}\) in real time. \\
The remaining diagonal terms \(q_{b_\omega}^{(i)}\) and \(q_{b_a}^{(i)}\) for \(i=1,2,3\) which govern the process noise associated with the gyroscope and accelerometer bias dynamics, are not learned by the network. Instead, they are specified a priori using conventional sensor calibration procedures and remain fixed during filter operation. These terms occupy the diagonal entries of \(\mathbf{Q}_{\mathrm{NN}}\) corresponding to the bias states in the 15-dimensional state vector. \\
\begin{table}[t]
\centering
\caption{Architecture of the proposed neural network for process noise estimation. 
The table reports the layer type, output dimensionality, and operations. 
Here, Conv1D denotes one-dimensional convolution, $k$ is the kernel size, and BN indicates batch normalization.}
\setlength{\tabcolsep}{4pt}
\renewcommand{\arraystretch}{1.1}
\begin{tabular}{l c l}
\hline
Layer & Dim. & Details \\
\hline
Conv1D & 32  & $k=5$, BN, LeakyReLU \\
Conv1D & 64  & $k=5$, BN, LeakyReLU \\
Conv1D & 128 & $k=5$, BN, LeakyReLU \\
Mean Pool & 128 & Temporal average \\
FC & 64 & LeakyReLU, Dropout \\
FC & 64 & LeakyReLU, Dropout \\
FC & 6 & Softplus \\
\hline
\end{tabular}
\label{fig:neuralnetwork1}
\end{table}
Our proposed neural-aided adaptive innovation-based Invariant Kalman filter is summarized in Algorithm \ref{alg:adaptive-filter}.
\begin{algorithm}[t]
\caption{Neural-Aided Adaptive Innovation-Based Invariant Kalman Filter}
\label{alg:adaptive-filter}
\begin{algorithmic}[1]
\Require Measurement covariance $N$,process noise covariance $Q_{\mathrm{base}}$, initial state $x_0$, covariance $P_0$
\State Initialize $x \gets x_0$, $P \gets P_0$
\State Initialize buffers $\mathcal{B}_{IMU}$ (size $M_{IMU}$) and $\mathcal{B}_{DVL}$ (size $M_{DVL}$)

\For{each measurement}

    \If{IMU measurement}
        \State Append measurement to $\mathcal{B}_{IMU}$
        \If{$\mathcal{B}_{IMU}$ is full}
            \State $\alpha \gets \mathrm{NeuralNetwork}(\mathcal{B}_{IMU})$ \eqref{neural-network-input}
            \State $Q \gets \lambda G\,\mathrm{diag}(\alpha,q_{b_\omega},q_{b_a})G^\top + (1-\lambda)Q_{innov}$  \eqref{hybrid-adaptive-Q}
        \Else
            \State $Q \gets GQ_{\mathrm{base}}G^\top$ \eqref{G-IKF}
        \EndIf
        \State $(x,P) \gets \mathrm{propagate}(x,P,Q)$ \eqref{F-IKF}
    \EndIf

    \If{DVL measurement}
        \State $(\delta x,P,K) \gets \mathrm{update\_error}(r,P,N)$
        \State $r \gets \mathrm{innov}(y,x)$ \eqref{innov-ikf}
        \State $\nu \gets K r$
        \State Append $\nu$ to $\mathcal{B}_{DVL}$
        \If{$\mathcal{B}_{DVL}$ is full}
            \State $Q_{innov} \gets \frac{1}{M_{DVL}}\sum \nu\nu^\top$ \eqref{adaptive-Q}
        \EndIf
        \State $x \gets \exp(\delta x^\wedge)x$ \eqref{correction-on-IKF}
    \EndIf

\EndFor
\State \Return $x,P$
\end{algorithmic}
\end{algorithm}
\subsection{Justification of the adaptive approach in the invariant setting}
\label{sec:org5828d99}
\begin{figure}
\centering
\includegraphics[width=\columnwidth]{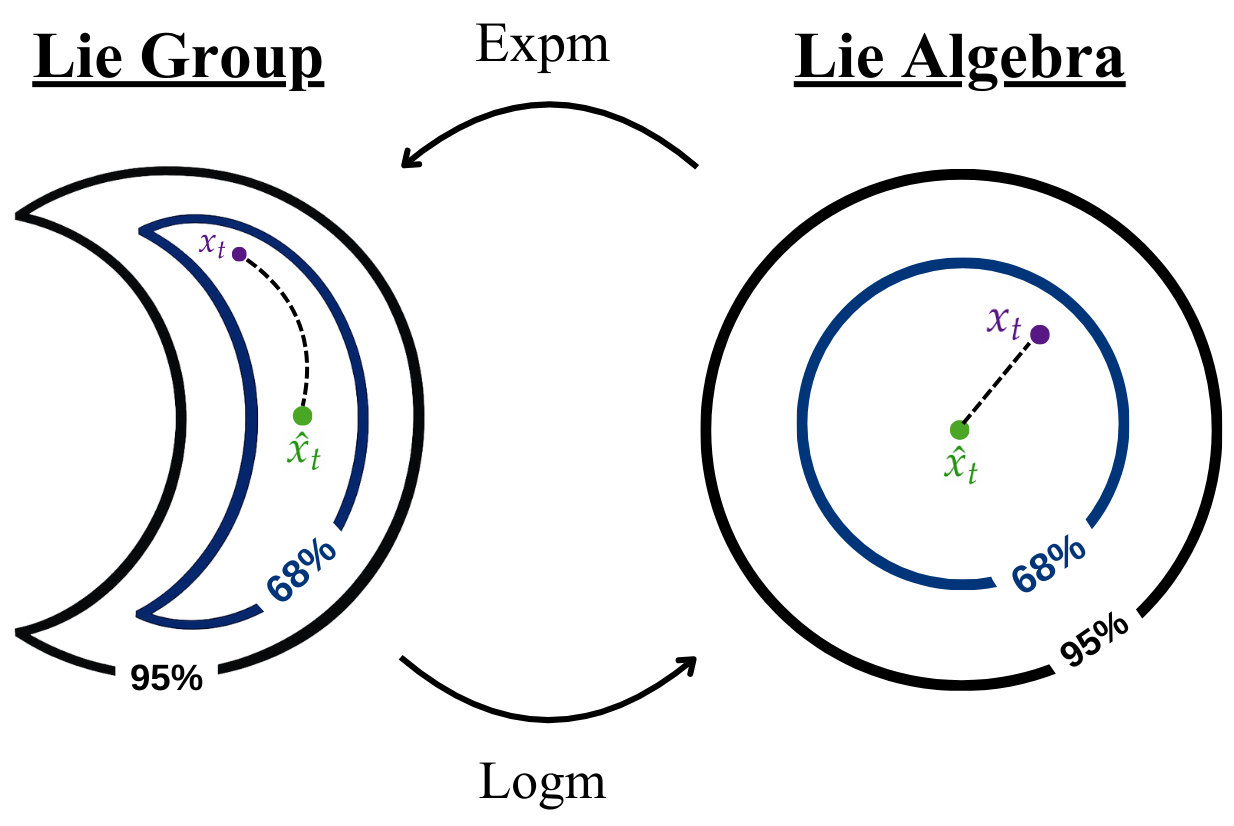}
\caption{Modeling the innovation as a Gaussian distribution in the Lie algebra results in a warped, banana-shaped distribution on the Lie group manifold. On the left, the Lie group representation displays the prior estimated state ($\hat{x}_t$, green) and the posterior estimated state ($x_t$, purple). On the right, the corresponding Lie algebra representation illustrates the same prior and posterior states, where the 68\% and 95\% uncertainty contours form standard concentric circles.}
\label{fig:innovation-lie-png}
\end{figure}

We assume that the invariant innovation projected into the Lie algebra (\ref{innov-equ}) follows a Gaussian distribution. Under this assumption, the corresponding innovation on the Lie group manifold (\ref{innov-eskf}) is not strictly Gaussian; rather, it takes on a warped geometry, as illustrated in Figure \ref{fig:innovation-lie-png}. \\
To leverage classical adaptive filtering techniques within this invariant framework, let \(\boldsymbol{r}_{innov} \sim \mathcal{N}(\boldsymbol{0},\mathbf{C}_{innov})\) denote the innovation as defined in (\ref{innov-ikf}). Following the methodology of, the state error \(\eta_{t}=\mathcal{\hat{X}}\mathcal{X}^{-1}_{t}\) on the Lie group is approximately distributed according to the Gaussian density of its Lie algebra equivalent, \(\boldsymbol{\zeta}\):
\begin{equation}
\begin{aligned}
\mathbb{P}(\eta_{t} \in \exp([\boldsymbol{\zeta},\boldsymbol{\zeta}+\boldsymbol{d\zeta}])) 
&\approx \frac{1}{(2\pi)^{M/2}|\mathbf{P}_{t}|^{1/2}}  \\
&\quad \times \exp\!\left(-\frac{1}{2}\boldsymbol{\zeta}^{t}
\mathbf{P}_{t}^{-1}\boldsymbol{\zeta}\right)d\boldsymbol{\zeta}
\end{aligned}
\end{equation}
where \(M\) is the number of estimated states in the invariant Kalman filter, \(\mathbf{P}_{t}\) is the theoretical state covariance matrix, and \([\boldsymbol{\zeta},\boldsymbol{\zeta}+\boldsymbol{d\zeta}]\subset \mathbb{R}^{M}\). \\
During the update step, the state correction applied to the Lie group is driven by the Kalman gain \(\mathbf{K}\). Consequently, the distribution of this correction step can be modeled as:
\begin{equation}
\begin{aligned}
&\mathbb{P}(\eta_{t} \in \exp([\mathbf{K}_{t}\boldsymbol{v_{t}},
\mathbf{K}_{t}\boldsymbol{v_{t}} + d(\mathbf{K}_{t}\boldsymbol{v_{t}})]))
\approx \\
&\frac{1}{(2\pi)^{M/2}|\mathbf{K}_{t}\mathbf{C}_{v_{k}}\mathbf{K}_{t}^{t}|^{1/2}} \\
&\times \exp\!\left(-\frac{1}{2}\boldsymbol{v_{t}}^{t}
(\mathbf{K}_{t}\mathbf{C}_{v_{k}}\mathbf{K}_{t}^{t})^{-1}
\boldsymbol{v_{t}}\right)d\boldsymbol{v_{t}}
\end{aligned}
\end{equation}

where \(M\in\mathbb{N}\) is the number of states, the innovation over lie algebra is \(\boldsymbol{v}_{t}:=\boldsymbol{r}_{innov}\) as in (\ref{innov-ikf}) and \(\mathbf{K_{t}}\) is the kalman gain of the current update. \\
Note that the classical Gaussian assumption of the innovation holds true only within the tangent space (Lie algebra), as demonstrated in Figure \ref{fig:innovation-lie-png}. \\
According to (\ref{correction-on-IKF}), each innovation evaluated on the Lie group can be exactly mapped to the corresponding state correction \(\delta\boldsymbol{x}\) in the Lie algebra via the logarithmic map: 
\begin{equation}
\log_m (\hat{\mathcal{X}}^{+} (\hat{\mathcal{X}}^{-})^{-1})^{\vee} = -\mathbf{K}\boldsymbol{r}_{innov}
\end{equation}
Using this relationship, we can calculate the empirical covariance of the state corrections in the Lie algebra over a sliding window of \(N\) epochs:
\begin{equation}
\mathbb{E}(\delta\boldsymbol{x} \delta\boldsymbol{x}^{t}) \approx \frac{1}{N} \sum_{n=1}^{N} \mathbf{K}_{n} \boldsymbol{r}_{innov_n} \boldsymbol{r}_{innov_n}^{t} \mathbf{K}_{n}^{t}
\end{equation}
By equating this empirical sample covariance with the theoretical filter covariance, the estimator for the process noise \(\hat{\mathbf{Q}}_k\) in the Lie algebra is given by \cite{mohamed1999adaptive}:
\begin{equation}
\hat{\mathbf{Q}}_{k} = \frac{1}{N} \sum_{j=k-N}^{k} \left( \delta\boldsymbol{x}_{j} \delta\boldsymbol{x}_{j}^{t} \right) + \mathbf{P}_{k|k} - \mathbf{\Phi}_{k-1} \mathbf{P}_{k-1|k-1} \mathbf{\Phi}_{k-1}^{t}
\end{equation}
Expanding \(\boldsymbol{r}_{innov}\) into its constituent terms, we arrive at our final adaptive law:
\begin{equation}
\hat{\mathbf{Q}}_k = \frac{1}{N} \sum_{j=k-N}^{k}\mathbf{K}_{j}\Pi(\mathcal{X}_{j}\boldsymbol{y}_{j}-\boldsymbol{b})[\Pi(\mathcal{X}_{j}\boldsymbol{y}_{j}-\boldsymbol{b})]^{t}\mathbf{K}_{j}^{t}
\end{equation}
where \(\Pi\) is the projection map as in (\ref{projection-map}), \(\mathcal{X}_{j}\) is the global state encoded as in (\ref{x-variable}), \(\boldsymbol{b}\in \mathbb{R}^{5}\) is defined at (\ref{b-vector}), and \(\boldsymbol{y}_{j}\in \mathbb{R}^{3}\) is the measurement defined at (\ref{new-measurement-y}).
\subsection{Sim-to-Real Training Perspective}
\label{sec:simulation}
The proposed training strategy follows a sim2real paradigm. Rather than attempting
to precisely model a specific real-world sensor, the neural network is trained on a wide
distribution of simulated trajectories, initial conditions, and noise levels. This
diversity encourages the network to learn invariant relationships between inertial
signal statistics and appropriate process noise parameters, which are expected to
generalize beyond the simulated domain. \\
Because underwater navigation experiments are costly and annotated datasets are scarce, the noise-estimation network is trained open-loop using simulation, where motion excitation and sensor-noise statistics are fully controlled. \\ 
To generate our simulative dataset, a total of 12 trajectory families are generated. There, each trajectory segment has a duration of
\[
T_{\text{traj}} = 60 \text{ s}
\]
with sampling time
\[
\Delta t = 0.01 \text{ s},
\]
resulting in
\[
N = \frac{T_{\text{traj}}}{\Delta t}  = 6000
\]
time steps per trajectory realization.
For each trajectory, four distinct IMU noise configurations are simulated. Each configuration
consists of a paired accelerometer and gyroscope noise level
\((\sigma_a, \sigma_\omega)\), where the pairs are defined as
\begin{equation}
\label{acc-noise-level}
\sigma_a \in
\{5 \times 10^{-1},\, 10^{-1},\, 5 \times 10^{-2},\, 10^{-2}\} \left[ \frac{m}{s^{2}\, \sqrt{Hz}} \right]
\end{equation}
\begin{equation}
\label{gyro-noise-level}
\sigma_\omega \in
\{10^{-4},\, 10^{-5},\, 10^{-6},\, 10^{-7}\} \left[\frac{rad}{s\, \sqrt{Hz}} \right]
\end{equation}
The accelerometer and gyroscope noise parameters are indexed together,
resulting in four paired IMU noise regimes per trajectory.
Thus, the total number of trajectory realizations is
\[
N_{\text{realizations}} = 12 \times 4 = 48.
\]
The total simulated time used for training is therefore
\[
T_{\text{total}} =
48 \times 60
= 2880 \text{ [s]}
= 48 \text{ minutes}
\]
All of the trajectories are generated in the ECEF frame with initial position of
\begin{equation}
\mathbf{p}_0 =
\begin{bmatrix}
4399229.20 \\
3068308.93 \\
3439906.25
\end{bmatrix}
[m]
\end{equation}
For each trajectory instance, the initial velocity \(\mathbf{v}_0\) is randomized and the body-to-ECEF orientation is initialized by a random \(\mathbf{R}_0\). The orientation is held constant over the 60 s segment:
\[
\mathbf{R}(t) \equiv \mathbf{R}_0
\]
The position is obtained by discrete-time integration of the velocity:
\[
\mathbf{p}_{k+1} = \mathbf{p}_{k} + \mathbf{v}_{k} \Delta t
\]
Kinematic motion families are simulated to span stationary, rectilinear, oscillatory, and turning maneuvers. Let \(\hat{\mathbf{d}}\) be the unit direction derived from \(\mathbf{v}_0\) and \(\hat{\mathbf{w}}\) a perpendicular unit vector. A nominal speed \(v_b = 5\,\text{m/s}\) is used.
\begin{itemize}
\item \textbf{Stationary}:
\[
  \mathbf{v}(t) = \mathbf{0}
  \]

\item \textbf{Straight (constant)}:
\[
  \mathbf{v}(t) = v_b \hat{\mathbf{d}}
  \]

\item \textbf{Straight (accelerating)}:
\[
  \mathbf{v}(t) = (v_b + a t)\hat{\mathbf{d}}, \quad a = 0.1\,\text{m/s}^2
  \]

\item \textbf{Straight (decelerating)}:
\[
  \mathbf{v}(t) = v_b \left(1 - \frac{t}{T_{\text{total}}}\right) \hat{\mathbf{d}}
  \]

\item \textbf{Oscillatory speed}:
\[
  \mathbf{v}(t) =
  v_b \left(1 + 0.5 \sin\left(\frac{2\pi t}{T_{\text{total}}}\right)\right)\hat{\mathbf{d}}
  \]

\item \textbf{Back-and-forth}:
\[
  \mathbf{v}(t) =
  v_b \, \mathrm{sign}\!\left(\sin\left(\frac{2\pi t}{T_p}\right)\right)\hat{\mathbf{d}}
  \quad T_p = 10\,\text{s}
  \]

\item \textbf{Vertical oscillation}:
\[
  \mathbf{v}(t) =
  2 \sin\left(\frac{2\pi t}{T_{\text{total}}}\right)\hat{\mathbf{u}}
  \quad \hat{\mathbf{u}} = \frac{\mathbf{p}_0}{\|\mathbf{p}_0\|}
  \]

\item \textbf{Spiral-like drift}:
\[
  \mathbf{v}(t) =
  \mathbf{v}_0 + \sin\left(\frac{4\pi t}{T_{\text{total}}}\right)\hat{\mathbf{r}}
  \]
where \(\hat{\mathbf{r}}\) is a random unit vector.

\item \textbf{Velocity random walk (clipped)}:
\[
  \mathbf{v}_{k} = \mathbf{v}_{k-1} + \delta\mathbf{v}_k
  \]
with bounded \(\|\mathbf{v}_k\|\).

\item \textbf{Lissajous-like}:
\end{itemize}
\begin{equation}
\begin{aligned}
v_x &= v_b \sin\!\left(\frac{2\pi t}{T_{\mathrm{total}}}\right) \\
v_y &= v_b \sin\!\left(\frac{4\pi t}{T_{\mathrm{total}}}\right) \\
v_z &=     \sin\!\left(\frac{6\pi t}{T_{\mathrm{total}}}\right)
\end{aligned}
\end{equation}

\begin{itemize}
\item \textbf{Circular}:
\[
  \mathbf{v}(t) =
  v_b \left(\cos(\omega t)\hat{\mathbf{d}} +
  \sin(\omega t)\hat{\mathbf{w}}\right)
  \quad \omega = \frac{2\pi}{T_{\text{total}}}
  \]

\item \textbf{Sinusoidal path}:
\[
  \mathbf{v}(t) =
  v_b \hat{\mathbf{d}} +
  \frac{v_b}{2} \sin(\omega t)\hat{\mathbf{w}}
  \]
\end{itemize}
Given \(\mathbf{p}^e(t)\), \(\mathbf{v}^e(t)\), and \(\mathbf{R}_b^e(t)\), the specific force vector is computed by:
\begin{equation}
\mathbf{f}^b =
(\mathbf{R}_b^e)^\top
\left(
\dot{\mathbf{v}}^e -
\mathbf{g}^e(\mathbf{p}^e) +
\boldsymbol{\Omega}_{ie}^e \boldsymbol{\Omega}_{ie}^e \mathbf{p}^e +
2 \boldsymbol{\Omega}_{ie}^e \mathbf{v}^e
\right)
\end{equation}
with
\[
\dot{\mathbf{v}}^e(t_k) \approx
\frac{\mathbf{v}^e_{k+1} - \mathbf{v}^e_k}{\Delta t}
\]
The angular velocity vector is
\[
\boldsymbol{\omega}_{ib}^b =
(\mathbf{R}_b^e)^\top \boldsymbol{\omega}_{ie}^e
\]
Lastly, the noisy IMU measurements are generated as
\[
\tilde{\mathbf{f}}^b = \mathbf{f}^b + \mathbf{n}_a \quad
\tilde{\boldsymbol{\omega}}^b =
\boldsymbol{\omega}_{ib}^b + \mathbf{n}_\omega
\]
where 
\begin{equation}
\mathbf{n}_a \sim \mathcal{N}(\mathbf{0}, \sigma_a^2 \mathbf{I}_3), \quad
\end{equation}
and  
\begin{equation}
\mathbf{n}_\omega \sim \mathcal{N}(\mathbf{0}, \sigma_\omega^2 \mathbf{I}_3)
\end{equation}
\subsection{Training}
\label{sec:org90126a3}
The process-noise covariance model is trained in a sim2real framework
using two alternative loss functions in separate experiments: Mean squared
error (MSE) and Huber loss. Let \(\hat{\mathbf{q}}_{\text{NN}}^{(i)}\) and
\(\mathbf{q}_{\text{NN}}^{(i)}\) denote the estimated and Ground truth (GT) process-noise
parameter vectors for the \(i\) -th training sample, and define the element-wise
error \(\mathbf{e}^{(i)} = \hat{\mathbf{q}}_{\text{NN}}^{(i)} -
\mathbf{q}_{\text{NN}}^{(i)} \in \mathbb{R}^d\).

When MSE is used for training, the loss is defined as:
\begin{equation}
\label{mse-loss}
\mathcal{L}_{\mathrm{MSE}}
= \frac{1}{N} \sum_{i=1}^{N} \left\| \mathbf{e}^{(i)} \right\|_2^2
\end{equation}
where \(N\) denotes the number of training samples. \\
Alternatively, in experiments using a more robust objective, the Huber loss is
adopted. For an element-wise error \(e\) , the Huber penalty is defined as
\begin{equation}
\label{huber-loss}
\ell_{\delta}(e) =
\begin{cases}
\frac{1}{2} e^2 & |e| < \delta \\
\delta\left(|e| - \frac{1}{2}\delta\right) & \text{otherwise}
\end{cases}
\end{equation}
where \(\delta>0\) is a threshold that determines the transition between the
quadratic and linear regimes. The total Huber loss is obtained by averaging over
samples and dimensions,
\[
\mathcal{L}_{\mathrm{Huber}}
= \frac{1}{Nd} \sum_{i=1}^{N}\sum_{j=1}^{d}
\ell_{\delta}\!\left(e^{(i)}_j\right)
\]
Regardless of the loss function used, the network parameters are optimized by minimizing the selected loss
function using stochastic gradient-based optimization. Let
\(\boldsymbol{\theta}\) denote the trainable parameters of the network.
The training objective is formulated as:
\[
\boldsymbol{\theta}^\ast
=
\arg\min_{\boldsymbol{\theta}} \mathcal{L}(\boldsymbol{\theta})
\]
where \(\mathcal{L}\) corresponds to either the MSE or Huber loss defined
above. Gradients of the loss with respect to the network parameters are
computed via backpropagation and evaluated over mini-batches of training
samples. Training is performed for 40 epochs. \\
Parameter updates are carried out using the Adam optimizer with a
learning rate of \(10^{-3}\). To reduce overfitting
and improve generalization, dropout with a probability of 0.2 is applied
during training. the default Adam momentum
coefficients are used.
\section{Analysis and Results}
\label{sec:orgc0ce113}
\subsection{Real-World Test Dataset}
\label{sec:akit-validation}
To examine our proposed approach on real recorded data, we employ the A-KIT dataset \cite{cohen2025adaptive}, publicly available at \href{https://github.com/ansfl/A-KIT}{A-KIT} . This dataset was collected in a real underwater environment and was
not used during training, thereby providing an independent benchmark for assessing
generalization beyond simulation. \\
The A-KIT dataset was recorded during sea trials conducted in the Mediterranean Sea
near the shore of Haifa, Israel, using the Snapir AUV.
The platform is equipped with a high-performance fiber-optic gyroscope 
inertial navigation system and a DVL, providing accurate
velocity measurements with a standard deviation of 0.02\textasciitilde{}m/s. The inertial navigation
system operates at 100\textasciitilde{}Hz, while the DVL measurements are available at 1\textasciitilde{}Hz. The dataset contains approximately seven hours of
data encompassing a wide range of maneuvers, depths, and vehicle speeds. GT
navigation estimates are provided by Delph INS post-processing software, which
generates a refined navigation solution based on the onboard inertial sensors and
external aiding. \\
In the original A-KIT study, the dataset comprised twelve distinct inertial and DVL data segments, each 400 seconds in duration, resulting in a total maneuvering time of 80 minutes. These segments were carefully selected to capture a broad range of vehicle dynamic behaviors. In contrast, the entire A-KIT dataset (twelve 400-second segments totaling 80 minutes) is used exclusively for testing and benchmarking against other navigation filters. The neural network for process-noise estimation is trained solely on simulated data using the sim-to-real strategy described in Section \ref{sec:simulation}.
\subsection{Comparison with Existing Methods}
\label{sec:org0e62773}
To assess the performance and relevance of the proposed adaptive invariant filtering approach, we compare it against representative model-based and learning-based methods that reflect current practice in adaptive Kalman filtering and neural innovation modeling for navigation.
\subsubsection{\textbf{Model-based adaptive EKF (AEKF)}}
\label{sec:org524d5eb}
The first baseline is a classical innovation-based AEKF, following the formulation in \cite{mohamed1999adaptive}. In this approach, the process noise covariance is adapted online based on statistics of the Euclidean innovation, defined as the difference between the measured and predicted outputs. This method has been widely used in navigation and robotics due to its simplicity, interpretability, and clear probabilistic foundations. However, it relies on linearized dynamics and an additive error definition, which may be inconsistent with the underlying geometry of nonlinear state spaces encountered in inertial navigation.
\subsubsection{\textbf{Model-based adaptive right-invariant Kalman filter (AR-IKF)}}
\label{sec:org1cb3b6b}
As a second baseline, we consider the model-based adaptive right-IKF. In contrast to the EKF, this filter is formulated on a Lie group and employs an invariant error definition consistent with the system's geometric structure. The innovation is defined through a projection onto the Lie algebra rather than via a Euclidean difference. While this formulation provides improved consistency with nonlinear dynamics, the adaptation of process noise remains purely model-driven and does not leverage data-driven representations of measurement residuals.
\subsubsection{\textbf{Neural innovation models within the EKF (NN-AEKF)}}
\label{sec:orgb6c5e0d}
\cite{or2022hybrid_ins_dvl}
To investigate the role of learning in classical filtering, we also evaluate neural innovation models embedded within the standard EKF framework. In these approaches, a neural network is trained to model or shape the innovation, effectively learning a data-driven correction to the measurement residual. We consider different training configurations, including single-step prediction (MSE step 1), multi-step prediction (step-10 MSE), and a robust variant using a Huber loss. These methods represent a recent trend in combining Kalman filtering with machine learning but remain tied to the Euclidean innovation structure of the EKF. \\
Together, these baselines span classical model-based adaptive filtering, geometry-aware invariant filtering, and recent neural innovation modeling techniques, providing a comprehensive basis for evaluating the proposed method.
\subsection{Performance Metrics}
\label{sec:org8997b39}
For performance evaluation, the root mean squared error (RMSE) of the position
estimate is used. Given the GT position \(\mathbf{p}(k)\) and its estimate
\(\hat{\mathbf{p}}(k)\) , the RMSE over \(N\) time steps is defined as:
\begin{equation}
\mathrm{RMSE}_{\mathbf{p}}
=
\sqrt{\frac{1}{N} \sum_{k=1}^{N}
\left\| \hat{\mathbf{p}}(k) - \mathbf{p}(k) \right\|_2^2 }
\end{equation}
\subsection{Results}
\label{sec:org27ceff8}
The neural adaptive innovation models were trained exclusively on simulated data, as described in Section \ref{sec:simulation}. In total, approximately 48 minutes of simulated data were used for training. Evaluation was performed on the A-KIT dataset, introduced in Section \ref{sec:akit-validation}. Importantly, none of the A-KIT trajectories were used during training. Instead, the A-KIT dataset serves exclusively for testing and benchmarking the different navigation filters. The evaluation set consists of 12 trajectories with a total duration of approximately 80 minutes, providing a wide range of motion conditions for assessing navigation performance. \\
For GT, we used the navigation solution provided by Delph INS, the post-processing software of iXblue's INS-based subsea navigation system. To create the unit under test, white Gaussian noise was added to the inertial and velocity measurements with the following power spectral densities: accelerometer noise of \(0.4 \left[\frac{m}{s^{2}\sqrt{Hz}}\right]\), gyroscope noise of \(10^{-4} \left[\frac{rad}{s\sqrt{Hz}}\right]\), and DVL noise of \(5\times 10^{-2} \left[\frac{rad}{s\sqrt{Hz}}\right]\) .

Table \ref{tab:unified-performance} compares the baseline approaches to NN-AEKF and our NN-AR-IKF in three setups: (1) S1-MSE, (2) S10-MSE, and (3) S10-Huber. These configurations correspond to different training strategies used to generate the neural network training samples. \\
In this context, S1-MSE denotes the case where training samples are generated at the inertial sensors sampling rate. Each sample consists of a fixed window of 100 consecutive inertial measurements and the network is trained using a standard MSE loss (\ref{mse-loss}). 
In contrast, S10-MSE refers to a subsampled training strategy in which consecutive training samples are generated every 10 inertial measurements. This reduces the temporal correlation between samples while maintaining the same 100-sample inertial window as input to the neural network. The S10-MSE configuration therefore uses the same loss function as S1 but with a step size of 10. 
Finally, the S10-Huber configuration follows the same sampling strategy as S10 but replaces the MSE loss (\ref{mse-loss}) with the Huber loss (\ref{huber-loss}) in order to increase robustness to outliers in the innovation prediction. \\
The results clearly demonstrate that embedding neural innovation
modeling within a geometrically consistent invariant filtering
framework yields substantial performance gains. While neural
enhancement produces only marginal improvement for the classical
EKF, its integration within the right-invariant Kalman filter
consistently reduces the estimation error across trajectories.
Notably, the proposed NN-AR-IKF achieves the lowest mean RMSE,
corresponding to an improvement of 17.1\% relative
to the AEKF baseline and 11.3\% relative to the NN-AEKF method. \\
These findings indicate that geometric invariance is not merely a
theoretical property but provides a structurally superior foundation
for learning-based innovation modeling. The consistent improvement
observed across real-world trajectories confirms the effectiveness
and robustness of the proposed neural-invariant formulation.
Furthermore, the fact that the neural network was trained solely on
simulated data yet demonstrates significant performance gains on
the real-world A-KIT dataset indicates that the proposed approach
successfully achieves a sim2real transfer.
This result highlights the practical applicability of the method in
domains such as underwater navigation, where collecting large-scale
labeled real-world training data is challenging.
\begin{table*}[t]
\centering
\caption{Position RMSE [m] on the A-KIT dataset. Comparison between classical and neural-enhanced EKF and right-invariant Kalman filter variants.}
\label{tab:unified-performance}
\setlength{\tabcolsep}{4pt}
\begin{tabular}{c|cc|ccc|ccc}
\hline
\textbf{Traj} 
& \textbf{AEKF} 
& \textbf{AR-IKF} 
& \multicolumn{3}{c|}{\textbf{NN-AEKF Variants}} 
& \multicolumn{3}{c}{\textbf{NN-AR-IKF Variants (ours)}} \\

\cline{4-9}

&  &  
& S1 (MSE) 
& S10 (MSE) 
& S10+Huber 
& S1 (MSE) 
& S10 (MSE) 
& S10+Huber \\

\hline
1  & 4.5 & 4.3 & 4.3 & 5.3 & 5.3 & 4.6 & 4.7 & 4.5 \\
2  & 3.8 & 3.5 & 3.8 & 3.5 & 3.7 & 2.9 & 2.9 & 3.0 \\
3  & 3.9 & 3.7 & 4.1 & 4.2 & 4.1 & 3.3 & 3.6 & 3.5 \\
4  & 5.1 & 4.6 & 4.5 & 5.0 & 5.6 & 3.1 & 3.8 & 4.0 \\
5  & 2.7 & 2.8 & 2.9 & 2.8 & 2.7 & 2.7 & 2.6 & 2.5 \\
6  & 40.0 & 50.4 & 37.1 & 38.8 & 38.3 & 36.0 & 38.4 & 37.8 \\
7  & 3.5 & 3.2 & 3.3 & 3.1 & 3.2 & 2.8 & 2.8 & 2.9 \\
8  & 3.6 & 2.5 & 3.9 & 3.4 & 4.1 & 4.4 & 2.1 & 2.4 \\
9  & 12.6 & 8.0 & 10.2 & 12.0 & 12.1 & 7.4 & 6.7 & 6.9 \\
10 & 5.3 & 5.1 & 5.6 & 5.3 & 5.7 & 4.3 & 4.5 & 4.6 \\
11 & 3.5 & 2.5 & 3.9 & 3.7 & 3.7 & 2.0 & 2.1 & 2.2 \\
12 & 2.2 & 2.1 & 2.3 & 2.3 & 2.5 & 2.2 & 2.2 & 2.2 \\
\hline
\textbf{Mean}
& 7.6 
& 7.7 
& 7.1 
& 7.5 
& 7.6 
& \textbf{6.3} 
& 6.4 
& 6.4 \\
\hline
\end{tabular}
\end{table*}

\section{Conclusions}
\label{sec:org5d74686}  
This paper addresses the incompatibility between traditional innovation-based process noise adaptation methods and the IKF framework. Classical innovation-based approach assumes Euclidean distance innovation and do not naturally align with the Lie-algebra error dynamics that govern invariant filtering. As a result, directly applying conventional adaptive covariance tuning within the IKF may compromise consistency or fail to fully exploit its structural properties.
To overcome this limitation, we derived a novel invariant adaptive innovation-based kalman filter. The proposed formulation respects the error dynamics defined on the Lie algebra and preserves the geometric structure of the IKF. Building upon this foundation, we introduced a lightweight neural network trained in a sim2real framework to estimate the process noise directly from inertial measurements. \\
Experimental evaluation on the A-KIT dataset demonstrates that the proposed adaptive IKF framework achieves a relative improvement of more than 17.1\% in position RMSE compared to an EKF employing an adaptive innovation-based algorithm. In addition to the quantitative improvement, the method exhibits enhanced innovation consistency and greater robustness across diverse motion regimes. The results show improved agreement between predicted covariance evolution and empirical estimation errors.
These findings confirm that adaptive process noise modeling can be naturally integrated into the invariant filtering framework when formulated consistently with its geometric structure. \\
The proposed approach nevertheless has certain limitations. Its effectiveness depends on the representativeness of the simulated training domain and the quality of domain randomization. Furthermore, although computationally lightweight, the neural network introduces additional complexity compared to fixed-noise implementations.\\
To conclude, from a practical perspective, the proposed framework provides a principled and scalable solution for adaptive uncertainty modeling in inertial fusion. More broadly, it demonstrates that learning-based components can be incorporated into geometry-aware estimation frameworks without compromising theoretical consistency. This work contributes toward bridging model-based filtering and neural-aided adaptation, supporting more reliable and robust autonomous navigation in real-world environments.
\printbibliography

\end{document}